\documentclass[11]{article}

\usepackage{amsfonts}
\usepackage{amsthm}
\usepackage{amsmath}
\usepackage{enumerate}
\usepackage{epic}
                                                             
\usepackage{graphicx}

\begin{document}




\null\vskip-24pt

\hfill KL-TH 01/02



\vskip0.3truecm

\begin{center}

\vskip 3truecm

{\Large\bf

\renewcommand{\thefootnote}{\fnsymbol{footnote}}
Composite fields, generalized hypergeometric functions and the $U(1)_Y$
symmetry in the AdS/CFT correspondence\footnote[1]{Talk given by W. R\"uhl at
the XXXVIIth Winter School on Theoretical Physics, Karpacz, Poland, February
2001} }\\
\renewcommand{\thefootnote}{\arabic{footnote}}
\vskip 1.5truecm


{\large\bf L. Hoffmann} \footnote{email:{\tt hoffmann@physik.uni-kl.de}}{\large\bf, T. Leonhardt} \footnote{email:{\tt tleon@physik.uni-kl.de}}{\large\bf, L. Mesref }\footnote{email:{\tt lmesref@physik.uni-kl.de}}{\large\bf, W.  R\" uhl} \footnote{email:{\tt ruehl@physik.uni-kl.de}}

\vskip 1truecm


{\it Department of Physics, Theoretical Physics\\

University of Kaiserslautern, Postfach 3049 \\

67653 Kaiserslautern, Germany}\\

\end{center}
\vskip 1truecm

\centerline{\bf Abstract}
We discuss the concept of composite fields in flat CFT as well as in the context of AdS/CFT. Furthermore we show how to represent Green functions using generalized hypergeometric functions and apply these techniques to four-point functions. Finally we prove an identity of $U(1)_Y$ symmetry for four-point functions.  

\newpage



\theoremstyle{plain}
\newtheorem{thm}{Theorem}
\theoremstyle{remark}
\newtheorem*{rem}{Remark}
\section{The concept of composite fields}

Consider a renormalizable CFT in $D$ spacetime dimensions defined by $k$
fundamental fields $\phi_1, \phi_2,\ldots, \phi_k$ of respective conformal
dimensions $\delta_1, \delta_2,\ldots, \delta_k$ and polynomial interactions.
We assume that this CFT is ``perturbative'': any $n$-point function can be
expanded into ``skeleton graphs'' with renormalized propagators
\begin{align}
\langle \phi_i (x) \phi_j (0) \rangle = \delta_{i,j} A_i (x^2)^{-\delta_i}.
\end{align}
These propagator normalizations $A_i$ can be cast on the vertex where
\begin{align}
\phi_{i_1} \phi_{i_2} \ldots  \phi_{i_f}
\end{align}
meet. Then this vertex obtains a coupling constant
\begin{align}
z(i_1,i_2,\ldots,i_f)^{\frac{1}{2}} = \biggl( \prod_{j=i_1}^{i_f} A_j \biggr)^{\frac{1}{2}}.
\end{align}
Assume that all vertices are also ``dressed''. Then the whole theory is
characterized by the dimensions $\delta_i$ and the couplings $z_i$. For each
propagator and each type of vertex there is a bootstrap equation which replace
the dynamical equations of the theory. They can be evaluated perturbatively.
In the limit of vanishing coupling, the CFT degenerates into a
\emph{generalized} free field theory which admits normal products. The
perturbative corrections of these normal product fields are the composite
fields. The composite fields must be conformal (or quasi-primary); this
eliminates ambiguities in this perturbative definition.

CFTs possess operator product expansions (OPEs): Each pair of conformal fields
can be expanded into (``blocks'' of) conformal fields and this expansion
converges if inserted into any $n$-point function in the maximal sphere
excluding the locations of the other $n\!-\!2$ fields. All conformal
fields appearing in such an OPE which are not fundamental are composite. A CFT
with $k$ fundamental fields may possess a sub~\!-~\!CFT (closed under OPE)
consisting solely of composite fields. E.g. a CFT with bosonic and fermionic
fields possesses a sub~\!-~\!CFT consisting of only the bosonic fields of the
CFT. If all fundamental fields are fermionic the sub~\!-~\!CFT consists of
bosonic composite fields. In a 1-particle reducible graph with exchange of a
fundamental field $\phi_i$ integration over the vertices attached to the
$\phi_i$-propagator gives two terms
\begin{align}
\sim F(\delta_i) (x^2)^{-\delta_i} + F(d-\delta_i)(x^2)^{-(d-\delta_i)},
\end{align}
which is the original propagator plus a ``shadow propagator'' . This symmetrization is the
consequence of conformal invariant integration. This shadow propagator can be
attributed to a shadow field $\phi_i^{(s)}$ of dimension $\delta_i^{(s)} = D
- \delta_i$, if Wightman positivity allows this, i.e. for a scalar field
besides the positivity condition for $\phi_i$
\begin{align}
\delta_i \geq \frac{D}{2} -1
\end{align}
also the positivity condition for $\phi_i^{(s)}$ has to be fulfilled:
\begin{align} \label{positiv}
D-\delta_i \geq \frac{D}{2} -1.
\end{align}
This means that $\delta_i$ must obey
\begin{align}
\frac{D}{2} +1 \geq \delta_i \geq \frac{D}{2} -1.
\end{align}
Let us assume that this is fulfilled. Then we can formulate
\begin{thm}\label{Thm1}
The CFT admits two different but equivalent representations. In the first we
have $\phi_i$ external legs, but no $\phi_i^{(s)}$ legs; in the second we
have $\phi_i^{(s)}$  external legs but no  $\phi_i$ legs.
\end{thm}
\begin{proof} The internal propagators are symmetric in $\phi_i$ and
$\phi_i^{(s)}$. An external $\phi_i$-leg goes into a $\phi_i^{(s)}$-leg by
amputation (their propagators are inverses as integral kernels), and vice
versa.
\end{proof}
\begin{rem}
 An artificial ``symmetrization'' by adding a  $\phi_i$-leg and a
$\phi_i^{(s)}$-leg is of no meaning.
\end{rem}
Turning our attention to OPEs we discover
\begin{thm} Fundamental fields appear with a direct and a shadow term,
composite fields only with a direct term, if OPEs are applied to n-point
functions.
\end{thm}
This is just everybody's experience with applications of OPEs. There is a complementary theorem:
\begin{thm}
The shadow field of a fundamental field cannot appear as a
composite field. The bootstrap equation for the propagator of a fundamental
field is equivalent to the absence of its shadow field as composite field.
\end{thm}
Theorem 3 is not so popular, therefore let us illustrate how it works in the
case of the critical non-linear $O(N)$ sigma model
\cite{lang} with the two fundamental fields
\begin{align} S_a (x):& \quad O(N) \;\textrm{vector, spacetime scalar} \\
\alpha (x): & \quad\textrm{the auxiliary field, } O(N)\; \textrm{and spacetime
scalar}
\end{align}
with propagators
\setlength{\unitlength}{1mm}
\begin{align}
\begin{picture}(115,0)
\put(30,0){$ \langle S_a (x) S_b (0) \rangle = A \,\delta_{ab} (x^2)^{-\delta_1} = $}
\put(78,0){\drawline(0,1)(8,1)}
\end{picture} \\
\begin{picture}(115,0)
\put(27,0){$\langle \alpha (x) \alpha (0) \rangle = B (x^2)^{-\delta_2} =
G(x) = $}
\put(80,0){\dashline{1}(0,1)(8,1)}
\end{picture}
\end{align}
with dimensions
\begin{equation}
\delta_1=\frac{1}{2}D-1+\mathcal{O}(\frac{1}{N}), \quad \delta_2=2+\mathcal{O}(\frac{1}{N}) 
\end{equation}
and interaction term
\begin{align}
z^{\frac{1}{2}} \int dx S_a (x)  S_a (x) \alpha (x), \qquad \textrm{with}
\quad z^{\frac{1}{2}} = A B^{\frac{1}{2}}.
\end{align}
The bootstrap equation for $G$ is
\setlength{\unitlength}{1mm}
\begin{align}\label{bootstrap}
\begin{picture}(110,0)
\put(30,0){$-G^{-1} = z$}
\put(52,1){\circle{8}}
\put(48,1){\circle*{2}}
\put(56,1){\circle*{2}}
\put(58,0){$+ z^2$}
\put(66,1){\circle*{2}}
\put(66,1){\drawline(0,0)(4,4)}
\put(66,1){\drawline(0,0)(4,-4)}
\put(66,1){\dashline{1}(4,4)(4,-4)}
\put(66,1){\drawline(4,4)(8,0)}
\put(66,1){\drawline(4,-4)(8,0)}
\put(74,1){\circle*{2}}
\put(76,0){$+ \cdots$}
\end{picture}
\end{align}
Consider the normal product with corrections
\begin{align}
T^{(0)} (x) = \sum_{a=1}^{N} : S_a (x)  S_a (x) : .
\end{align}
If this field exists, it should be the shadow field of $\alpha$. Its dimension
in $2<D<4$ is
\begin{align}
\textrm{dim}\, T^{(0)} = D - \textrm{dim}\,\alpha = D-2+\mathcal{O}(\frac{1}{N}).
\end{align}
Since the dimension of $ T^{(0)}$ fulfills the positivity condition
(\ref{positiv}) one would expect by theorem \ref{Thm1} that we could
reformulate the $O(N)$ sigma model in terms of $T^{(0)}$. The definition of $
T^{(0)}$ implies  that it should appear in the OPE of $S_a (x)  S_a (y)$.
Calculation of the 4-point function
\begin{align}
\langle S_a(x+\varepsilon_1) S_a(x-\varepsilon_1) S_b(y+\varepsilon_2)
S_b(y-\varepsilon_2) \rangle
\end{align}
in the limit $\varepsilon \rightarrow 0$ and subsequent decomposition into
1-P-reducible and -irreducible graphs gives
\begin{align}
\begin{picture}(115,0)
\put(20,0){$ \langle T^{(0)} (x) T^{(0)} (y) \rangle \,\sim $}
\put(52,1){\circle*{2}}
\put(56,1){\circle{8}}
\put(55,-0.4){F}
\put(60,1){\circle*{2}}
\put(62,0){$ +$}
\put(67,1){\circle*{2}}
\put(71,1){\circle{8}}
\put(70,-0.4){F}
\put(75,1){\circle*{2}}
\put(76,1){\dashline{1}(0,0)(6,0)}
\put(83,1){\circle*{2}}
\put(86,-0.4){F}
\put(87,1){\circle{8}}
\put(91,1){\circle*{2}}
\end{picture}
\end{align}
where the circles with the Fs are (up to renormalization) the r.h.s. of the
$\alpha$-bootstrap equation (\ref{bootstrap}). Call it $-\tilde G^{-1}$.
Then
\begin{align}
\langle  T^{(0)} (x)  T^{(0)} (y) \rangle \sim -  \tilde G^{-1} +   \tilde
G^{-1} G \tilde G^{-1} . \end{align}
The bootstrap equation says $G^{-1} = \tilde G^{-1}$, therefore
\begin{align}
\langle  T^{(0)} (x)  T^{(0)} (y) \rangle = 0.
\end{align}
On the other hand
\begin{align}
\langle  T^{(0)} (x)  T^{(0)} (y) \rangle = 0 \Longrightarrow    G^{-1} = \tilde G^{-1}.
\end{align}

Now we turn to the AdS/CFT correspondence and apply our theorems. Any field $\phi$ on AdS with mass
$m^2 \geq -\frac{D^2}{4}$ is mapped holographically  on a CFT field $\tilde
\phi$ with dimension
\begin{align}
\delta = \frac{D}{2} \pm \Bigl( \frac{D^2}{4} + m^2 \Bigr)^{\frac{1}{2}},
\end{align}
where the upper/lower sign corresponds to $\tilde \phi$/$\tilde \phi ^{(s)}$
respectively. In this context Aharony et al. \cite{aharony} considered
square integrability and boundary conditions. But we have
\begin{thm}
 Square integrability of an AdS wave function and Wightman
positivity of the corresponding CFT field are the same thing.
\end{thm}
Thus we arrive at the conclusion: Taking for a field always the upper sign or always the lower sign
(if Wightman positivity allows this) leads to two different but equivalent CFTs.

By evaluating AdS exchange graphs and analyzing the result one finds that
there are no shadow terms \cite{liu}. This leads us to
\begin{thm}
The holographic image of an AdS field theory is a CFT without
fundamental fields.
\end{thm}
In fact in the standard example of $\mathcal{N}=4$ SYM$_4$,
which is dual to type $I\!I\!B$ superstring theory on $\textrm{AdS}_5 \times
S^5$,
 the relevant coupling constant is `t Hooft's
\begin{align}
\lambda = g_{YM}^2 N .
\end{align}
It is a perturbative CFT  if $\lambda \rightarrow 0$ with (say) the gauge supermultiplett as
fundamental fields. They belong to the adjoint representation of the gauge group SU(N). The
holographic image possesses only gauge invariant fields which are therefore composed of at least two
fundamental fields with the adjoint representation matrices ``traced'' away.

But the content of Theorem 5 derives from the geometry of $\textrm{AdS}_5$ and not from gauge
invariance. In other examples compositeness and gauge invariance may not have the same effect.


\section{Generalized hypergeometric functions}
\setcounter{equation}{0}
A ``star graph'' (vertex) of $m$ scalar fields in AdS$_{D+1}$ conformal field theory 
§§§§§§§§§§§§§§§§§§§§§§§§§§§§§§§§§§§§§§§§§§§§§§§§§§§§§§§§§§§§§§§§§§§§§
\begin{figure}[htb]
\begin{centering}
\includegraphics[scale=0.45]{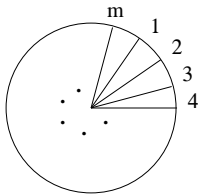}  
\caption{AdS$_{D+1}$ ``star graph'' (vertex)}
\end{centering}
\end{figure}
§§§§§§§§§§§§§§§§§§§§§§§§§§§§§§§§§§§§§§§§§§§§§§§§§§§§§§§§§§§§§§§§§§§§§

defines a Green function ($x_i \in {\bf{R}}_D = \partial\text{AdS}_{D+1}$)
\begin{equation} \label{2.1}
\Gamma(x_1,...,x_m)= \pi^{-\frac{D}{2}} \int_{y_0>0} \frac{d^{D+1}y}{y_0^{D+1}} \prod_{i=1}^m (\frac{y_0}{y_0^2 + (\vec{y}-\vec{x_i})^2})^{\alpha_i} 
\end{equation}
where$\{\alpha_i\}$ are field dimensions constrained only by 
\begin{equation} \label{2.2}
\alpha_i \geq \frac{1}{2} D -1. 
\end{equation} 
In conformal field theory on $\bf{R}_D$ such a vertex is given by
\begin{equation} \label{2.3}
G_m(x_1,...,x_m) = \pi^{-\frac{D}{2}} \int_{\bf{R}_D} d^Dy \prod_{i=i}^m ((y-x_i)^2)^{-\alpha_i} 
\end{equation}
with the conformal ``uniqueness'' condition
\begin{equation} \label{2.4}
\sum_{i=1}^m \alpha_i = D. 
\end{equation}
The usual parametric representation of $G_m$
\begin{equation} \label{2.5}
G_m(x_1,...,x_m) = [\prod_{i=1}^m \Gamma(\alpha_i)]^{-1}(\prod_{i=1}^m \int_0^\infty dt_i \, t_i^{\alpha_i-1}) \, T^{-\frac{D}{2}} e^{-\frac{1}{T} \sum_{i<j}t_it_j x_{ij}^2} 
\end{equation}
with
\begin{equation} \label{2.6}
T = \sum_{i=1}^m t_i.  
\end{equation}
allows to continue analytically in $\{\alpha_i\}$ and $D$ independently, such that one has the Green function
\begin{equation} \label{2.7}
G_m(D;\alpha_1,...,\alpha_m). 
\end{equation}
The AdS$_{D+1}$ function $\Gamma_m$ is \cite{hoff}
\begin{equation} \label{2.8}
\Gamma_m(D;\alpha_1,...,\alpha_m) = \frac{1}{2} \Gamma(\frac{1}{2}(\sum_{i=1}^m \alpha_i - D)) \, G_m(\sum_{i=1}^m \alpha_i;\alpha_1,...,\alpha_m). 
\end{equation}
All the Green functions $G_m$ have been represented in the form of a Mellin-Barnes integral and by expansion in generalized hypergeometric form by K. Symanzik \cite{symanzik}.
For $m=4$ and with the biharmonic ratios
\begin{equation} \label{2.9} 
u = \frac{x_{13}^2 x_{24}^2}{x_{12}^2 x_{34}^2}, \quad  v = \frac{x_{14}^2 x_{23}^2}{x_{12}^2 x_{34}^2} 
\end{equation}
we get
\begin{equation} \label{2.10}
G_4(x_1, x_2, x_3, x_4) = [\prod_{i=1}^4\Gamma(\alpha_i)]^{-1} (x_{12}^2)^{\alpha_4-\frac{1}{2}D} (x_{14}^2)^{\frac{1}{2}D-\alpha_1-\alpha_4} (x_{24}^2)^{\frac{1}{2}D-\alpha_2-\alpha_4} (x_{34}^2)^{-\alpha_3} F(u,v) 
\end{equation}
with
\begin{align} \label{2.11}
F(u,v) = & u^{-\frac{1}{2}(\alpha_1 + \alpha_3)} \sum_{n,m=0}^\infty \frac{u^n (1-v)^m}{n! m!} \{u^{\frac{1}{2}(\alpha_1 + \alpha_3)} \Gamma(\frac{1}{2}(\alpha_2+\alpha_4-\alpha_1-\alpha_3)) \notag \\
&\times \frac{\Gamma(\alpha_1+n) \Gamma(\frac{1}{2}D-\alpha_2+n) \Gamma(\alpha_3+n+m) \Gamma(\frac{1}{2}D-\alpha_4+n+m)}{\Gamma(\alpha_1+\alpha_3+2n+m)(\frac{1}{2}(\alpha_1+\alpha_3-\alpha_2-\alpha_4)+1)_n}+ u^{\frac{1}{2}(\alpha_2 + \alpha_4)} \notag \\ & \times\Gamma(\frac{1}{2}(\alpha_1+\alpha_3-\alpha_2-\alpha_4)) \frac{\Gamma(\alpha_4+n) \Gamma(\frac{1}{2}D-\alpha_3+n) \Gamma(\alpha_2+n+m) \Gamma(\frac{1}{2}D-\alpha_1+n+m)}{\Gamma(\alpha_2+\alpha_4+2n+m)(\frac{1}{2}(\alpha_2+\alpha_4-\alpha_1-\alpha_3)+1)_n}. 
\end{align}
This form converges in a complex neighborhood of 
\begin{equation} \label{2.12}
u = 0, \quad v=1 
\end{equation}
and is therefore suited for an operator product expansion (OPE) in the limit, ``$t$-channel''
\begin{equation} \label{2.13}
x_{13} \longrightarrow 0, \quad x_{24} \longrightarrow 0. 
\end{equation}  
The analytic continuation to the ``$s$-channel''
\begin{align}
&x_{12} \longrightarrow 0, \quad x_{34} \longrightarrow 0, \notag \\
&u \underset{g_{t \rightarrow s}}{\longrightarrow} u'= \frac{1}{u}, \notag \\
&v \underset{g_{t \rightarrow s}}{\longrightarrow} v'= \frac{v}{u}  
\end{align}
and to the ``$u$-channel''
\begin{align}
& x_{14} \longrightarrow 0, \quad x_{23} \longrightarrow 0, \notag\ \\
&u \underset{g_{t \rightarrow u}}{\longrightarrow} u''= v, \notag  \\
&v \underset{g_{t \rightarrow u}}{\longrightarrow} v''= u  \end{align}
can be obtained by two methods
\begin{enumerate}
\item using the symmetry group of the graphs \cite{hoff};
\item using the Kummer formulae for $_2F_1$-functions \cite{grad}.
\end{enumerate}

The AdS graphs at order $\mathcal{O}(\frac{1}{N^2})$ for dilaton-axion four-point functions with 
\begin{equation} \label{2.20}
m^2=0 \longrightarrow \Delta = 4 \quad (D = 4) 
\end{equation}
can all be represented by
\begin{equation} \label{2.21}
\Gamma_4(x_1,x_2,x_3,x_4)|_{\alpha_1=\alpha_3=\Delta, \alpha_2=\alpha_4=\Delta'; \Delta, \Delta'\in \bf{N}; \Delta' \geq \Delta} = \pi^{-\frac{1}{2}D} (x_{12}^2)^{-\Delta}(x_{24}^2)^{\Delta-\Delta'}(x_{34}^2)^{-\Delta} G_{\Delta \Delta'}(u, v) 
\end{equation}
with 
\begin{align} 
G_{\Delta \Delta'}(u, v) &= \frac{\pi^2}{2} \frac{\Gamma(\Delta+\Delta'-2)}{\Gamma(\Delta)^2 \Gamma(\Delta')^2} \sum_{m=0}^\infty \frac{(1-v)^m}{m!}\{\sum_{n=0}^{\Delta'-\Delta-1} \frac{(-1)^n u^n}{n!} (\Delta'-\Delta-n-1)! \notag \\ &\times \frac{\Gamma(\Delta+n)^2 \Gamma(\Delta+n+m)^2}{\Gamma(2\Delta+2n+m)} +\sum_{n=\Delta'-\Delta}^\infty \frac{(-1)^{\Delta'-\Delta}u^n}{n!(n-\Delta'+\Delta)!} \frac{\Gamma(\Delta+n)^2\Gamma(\Delta+n+m)^2}{\Gamma(2\Delta+2n+m)} \notag \\
&\times[-log\, u + \Psi(n- \Delta'+\Delta+1) \notag \\ &+ \Psi(n+1) -2\Psi(\Delta+n) +2\Psi(2\Delta+2n+m)-2\Psi(\Delta+n+m)]\}. 
\end{align} 
In the other channels one has
\begin{align}
G_{\Delta \Delta'}(u', v') =& \frac{\pi^2}{2} \frac{\Gamma(\Delta+\Delta'-2)}{\Gamma(\Delta)^2 \Gamma(\Delta')^2} u^{\Delta} \sum_{n,m=0}^\infty \frac{u^n (1-v)^m}{(n!)^2 m!} \notag \\
&\times \frac{\Gamma(\Delta+n) \Gamma(\Delta'+n)\Gamma(\Delta+n+m)\Gamma(\Delta'+n+m)}{\Gamma(\Delta+\Delta'+2n+m)} \{-log \, u +2 \Psi(n+1)-\Psi(\Delta+n)\notag \\&-\Psi(\Delta'+n)-\Psi(\Delta+n+m)-\Psi(\Delta'+n+m)+2\Psi(\Delta+\Delta'+2n+m)\} 
\end{align}
and 
\begin{align} 
G_{\Delta \Delta'}(u'', v'') =& \frac{\pi^2}{2} \frac{\Gamma(\Delta+\Delta'-2)}{\Gamma(\Delta)^2 \Gamma(\Delta')^2} v^{\Delta'-\Delta} \sum_{n,m=0}^\infty \frac{u^n (1-v)^m}{(n!)^2 m!} \notag \\
&\times \frac{\Gamma(\Delta'+n)^2 \Gamma(\Delta+n+m)^2}{\Gamma(\Delta+\Delta'+2n+m)} \{-log \, u +2 \Psi(n+1)-2\Psi(\Delta+n)-2\Psi(\Delta'+n+m) \notag \\
&+2\Psi(\Delta+\Delta'+2n+m)\}. 
\end{align}.

\section{$U(1)_Y$ symmetry}
\setcounter{equation}{0}
In the standard model of AdS/CFT correspondence
\begin{equation} \label{3.1}
AdS_5 \times S_5 \longrightarrow \text{SYM}_4 
\end{equation}
the $U(1)_Y$ symmetry carries over to the strong coupling domain of SYM$_4$ and makes predictions on three and four-point functions \cite{intriligator}. We will discuss here an identity of two four-point functions \cite{hoff}.
Let 
\begin{align}
\text{Dilaton}:& \quad \Phi(y) \longrightarrow \tilde{\Phi}(x) \sim Tr(F_{\mu \nu}(x) F_{\mu \nu}(x)) \notag \\
\text{Axion}:& \quad C(y) \longrightarrow \tilde{C}(x) \sim Tr(F_{\mu \nu}(x) \tilde{F}_{\mu \nu}(x))
\end{align}
and consider
\begin{equation}
\mathcal{O}_\tau(x) = \tilde{\Phi}(x) + i \tilde{C}(x). 
\end{equation}
This field has $U(1)_Y$ charge $q=-4$.
Therefore
\begin{equation} \label{a}
<\prod_{i=1}^4 \mathcal{O}_\tau(x)> = 0, 
\end{equation}
due to charge non-conservation ($\sum_i q_i \neq 0$).
Introducing the bilocal operators
\begin{align} 
&\Psi_{\mp}(x_1, x_3) = \frac{1}{\sqrt{2}}[\mp \tilde{\Phi}(x_1)  \tilde{\Phi}(x_3)+ \tilde{C}(x_1)  \tilde{C}(x_3)]  \\
&\Psi_0(x_1,x_3)=\frac{1}{\sqrt{2}}[\tilde{\Phi}(x_1)\tilde{C}(x_3)+\tilde{C}(x_1)\tilde{\Phi}(x_3)] 
\end{align}
which by OPE produce tensor fields of even rank with opposite parities, we obtain from (\ref{a})
\begin{equation} \label{b}
<\Psi_{-}(x_1,x_3) \Psi_{-}(x_2,x_4)> - <\Psi_{0}(x_1,x_3) \Psi_{0}(x_2,x_4)> = 0. 
\end{equation}
This relation can be worked out up to order $\mathcal{O}(\frac{1}{N^2})$ at present.
At leading order $\mathcal{O}(1)$, both four-point functions are
\begin{equation} 
(x_{12}^2x_{34}^2)^{-4}(1+v^{-4}) \quad \text{with} \quad v = \frac{x_{14}^2 x_{23}^2}{x_{12}^2 x_{34}^2}. 
\end{equation}
§§§§§§§§§§§§§§§§§§§§§§§§§§§§§§§§§§§§§§§§§§§§§§§§§§§§§§§§§§§§§§§§§§§§§§§
\begin{figure}[htb]
\begin{centering}
\includegraphics[scale=0.45]{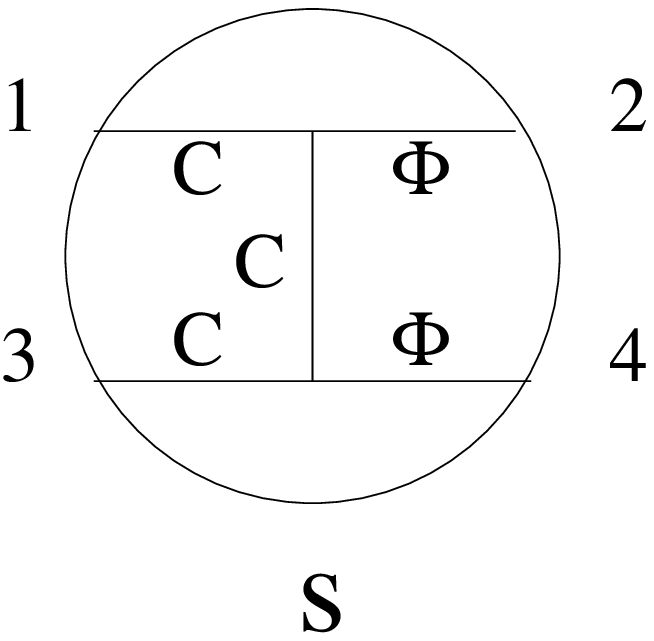}\qquad \includegraphics[scale=0.45]{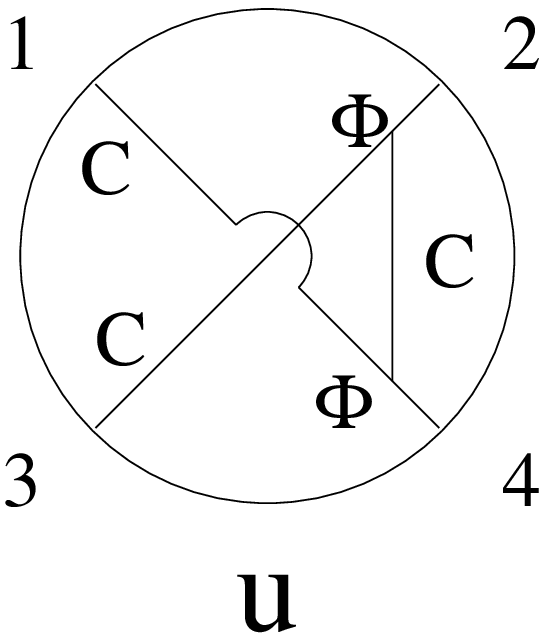} \qquad \includegraphics[scale=0.45]{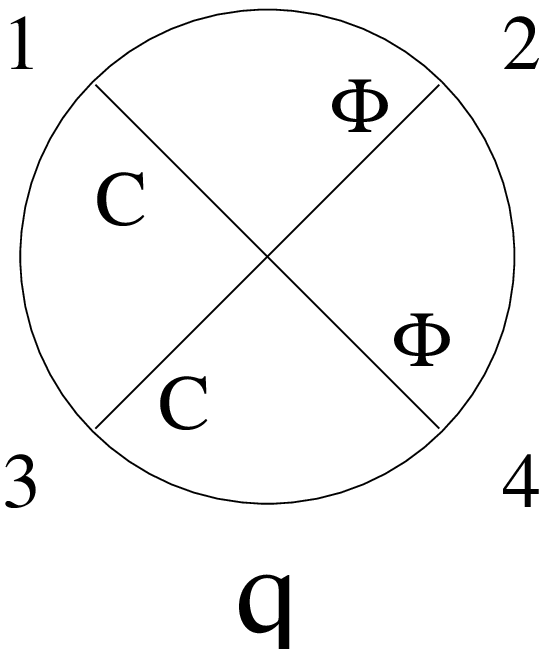}
\caption{Relevant graphs at order $\mathcal{O}(\frac{1}{N^2})$}
\end{centering}
\end{figure}
§§§§§§§§§§§§§§§§§§§§§§§§§§§§§§§§§§§§§§§§§§§§§§§§§§§§§§§§§§§§§§§§§§§§§§§§
At order $\mathcal{O}(\frac{1}{N^2})$ the relevant graphs are given by Figure 2, while the dilaton and graviton exchange graphs cancel in (\ref{b}).
The sum is denoted 
\begin{equation}
I_{(1,3)}^{(s+u+q)},
\end{equation}
where we indicate the position of the $C$-fields.
However,
\begin{equation}
I_{(1,3)}^{(s+u+q)} = I_{(2,4)}^{(s+u+q)}.
\end{equation}
Then the constraint (\ref{b}) is expressible as
\begin{equation} \label{c}
I_{(1,3)}^{(s+u+q)} + I_{(1,2)}^{(s+u+q)} +  I_{(1,4)}^{(s+u+q)} = 0.
\end{equation}
This is typically a constraint for a crossing invariant function.

Now we take from d'Hoker et al. \cite{dhoker}
\begin{equation}
I_{(1,3)}^{(s+u+q)} = 32 (\frac{6}{\pi^2})^4 (x_{12}^2x_{34}^2)^{-4} [2 G_{45}(u,v)-G_{44}(u,v)] 
\end{equation}
so that, with the analytic continuations in all channels, we get from (\ref{c})
\begin{align} \label{d}
G_{45}(u,v)+u^{-4}G_{45}(u',v')+G_{45}(u'',v'')=& \frac{1}{2}[G_{44}(u,v)+u^{-4}G_{44}(u',v')+G_{44}(u'',v'')]\notag \\=&\frac{3}{2}G_{44}(u,v) 
\end{align}
since $G_{44}$ is a star function with equal legs and therefore itself crossing invariant.

In fact the following identity can be proved
\begin{equation} \label{e}
G_{\Delta,\Delta+1}(u,v)+u^{-\Delta}G_{\Delta,\Delta+1}(u',v')+G_{\Delta,\Delta+1}(u'',v'') = \frac{2\Delta-2}{\Delta} G_{\Delta,\Delta}(u,v). 
\end{equation}
It is a recursion for the generalized hypergeometric function $G$.

For the proof of (\ref{e}) we do not use the explicit form of $G_{\Delta,\Delta'}$ given earlier, but make use of an ``$\epsilon$-trick''
\begin{align}
G_{\Delta,\Delta'}(u,v) = \underset{\epsilon \rightarrow 0}{lim}& \frac{\partial}{\partial\epsilon}[\frac{\pi^2}{2}\frac{\Gamma(\Delta+\Delta'-2)}{\Gamma(\Delta)^2 \Gamma(\Delta')^2} (-1)^{\Delta'-\Delta}\prod_{k=1}^{\Delta'-\Delta}(n-\Delta'+\Delta+k-\epsilon)\notag \\ &\times \sum_{m,n=0}^\infty \frac{u^{n-\epsilon}(1-v)^m}{\Gamma(n+1-\epsilon)^2 m!} \frac{\Gamma(\Delta+n-\epsilon)^2 \Gamma(\Delta+n+m-\epsilon)^2}{\Gamma(2\Delta+2n+m-2\epsilon)}],
\end{align}
 \begin{align}
u^{-\Delta} G_{\Delta,\Delta'}(u',v')=& \underset{\epsilon \rightarrow 0}{lim} \frac{\partial}{\partial\epsilon}[\frac{\pi^2}{2}\frac{\Gamma(\Delta+\Delta'-2)}{\Gamma(\Delta)^2 \Gamma(\Delta')^2} \sum_{m,n=0}^\infty \frac{u^{n-\epsilon}(1-v)^m}{\Gamma(n+1-\epsilon)^2 m!} \notag \\ &\times \frac{\Gamma(\Delta+n-\epsilon)\Gamma(\Delta'+n-\epsilon) \Gamma(\Delta+n+m-\epsilon) \Gamma(\Delta'+n+m-\epsilon)}{\Gamma(\Delta+\Delta'+2n+m-2\epsilon)}],
\end{align}
\begin{equation}
G_{\Delta,\Delta'}(u'',v'') = \underset{\epsilon \rightarrow 0}{lim} \frac{\partial}{\partial\epsilon}[\frac{\pi^2}{2}\frac{\Gamma(\Delta+\Delta'-2)}{\Gamma(\Delta)^2 \Gamma(\Delta')^2}  \sum_{m,n=0}^\infty \frac{u^{n-\epsilon}(1-v)^m}{\Gamma(n+1-\epsilon)^2 m!} \frac{\Gamma(\Delta'+n-\epsilon)^2 \Gamma(\Delta+n+m-\epsilon)^2}{\Gamma(\Delta+\Delta'+2n+m-2\epsilon)}]. 
\end{equation}
Since $\Delta'-\Delta \in \bf{N}_0$, we can sum the coefficients as
\begin{align}
\frac{\Gamma(\Delta+n-\epsilon)^2 \Gamma(\Delta+n+m-\epsilon)^2}{\Gamma(\Delta+\Delta'+2n+m-2\epsilon)} &\{(2\Delta+2n+m-2\epsilon)_{\Delta'-\Delta} \prod_{k=1}^{\Delta'-\Delta}(\Delta'-\Delta-n-k+\epsilon) \notag \\ &+(\Delta+n-\epsilon)^2_{\Delta'-\Delta}+(\Delta+n-\epsilon)_{\Delta'-\Delta}(\Delta+n+m-\epsilon)_{\Delta'-\Delta}\} 
\end{align}
For $\Delta' = \Delta+1$ we have
\begin{equation}
\{ ... \} = \Delta(2\Delta+ 2n+m-2\epsilon) 
\end{equation}
which leads to $G_{\Delta \Delta}$ by the formula given above. This completes the proof.

Let us quote some detailed results on the OPE of the bilocal fields $\Psi_{+,-,0}$ \cite{hoff}.
\begin{enumerate}
\item 
$\Psi_{+}(x_1,x_3)$ expands into the stress-energy tensor plus the towers of conformal fields:
tensor rank $l \in 2\bf{N}_0$,
parity $(+1)$ and
dimension 
\begin{equation}
\delta_{+}(l,t)=8+l+2t+\eta_{+}(l,t); \quad t \in {\bf{N}_0} (\text{tower label, twist}); \eta_{+}= \mathcal{O}(\frac{1}{N^2}) 
\end{equation}
\item
$\Psi_{-}(x_1,x_3)$ expands into conformal fields:
tensor rank $l \in 2\bf{N}_0$,
parity $(+1)$ and
dimension 
\begin{equation}
\delta_{-}(l,t)=8+l+2t+\eta_{-}(l,t); \quad t\in {\bf{N}_0}, \eta_{-}= \mathcal{O}(\frac{1}{N^2}) 
\end{equation}
\item 
$\Psi_{0}(x_1,x_3)$ expands into conformal fields:
tensor rank $l \in 2\bf{N}_0$,
parity $(-1)$ and
dimension 
\begin{equation}
\delta_{0}(l,t)=8+l+2t+\eta_{0}(l,t); \quad t\in {\bf{N}_0}, \eta_{0}= \mathcal{O}(\frac{1}{N^2}) 
\end{equation}
\end{enumerate}

For $t=0$ we obtain
\begin{equation}
\eta_{-}(l,0) = \eta_{0}(l,0) = -\frac{96}{N^2(l+1)(l+6)} 
\end{equation}
and the corresponding fusion constants squared
\begin{equation}
\epsilon_{-}(l,0) = \epsilon_{0}(l,0) = \frac{1}{18}(2l+7)(l+1)_6  
\end{equation}
The remaining quantities $\eta_{+}, \epsilon_{+}$ and $\eta_{-}=\eta_{0} (t \neq 0)$,  $\epsilon_{-}=\epsilon_{0} (t \neq 0)$ are easily accessible by using computer manipulations. One has to solve infinite linear equations with triangular matrices, where $t$ counts the off-diagonality of the matrix elements. The main labor resides in  calculating  the matrix elements in a generic form (no numbers).

\end{document}